# Performance of synthetic antiferromagnetic racetrack memory: domain wall vs skyrmion

R. Tomasello[1], V. Puliafito,[2] E. Martinez[3], A. Manchon[4], M. Ricci,[1] M. Carpentieri[5], and G. Finocchio[6*]

[1]Department of Engineering, Polo Scientifico Didattico di Terni, University of Perugia, strada di Pentima Bassa, I-50100 Terni, Italy

[2]Department of Engineering, University of Messina, C.da di Dio, I-98166, Messina, Italy.

[3]Department of Fisica Aplicada, Universidad de Salamanca, Plaza de los Caidos s/n, E-38008, Salamanca, Spain.

[4] Physical Science and Engineering Division (PSE), King Abdullah University of Science and Technology (KAUST),  Thuwal 23955-6900, Saudi Arabia

[5]Department of Electrical and Information Engineering, Politecnico di Bari, via E. Orabona 4, I-70125 Bari, Italy.

[6]Department of Mathematical and Computer Sciences, Physical Sciences and Earth Sciences, University of Messina, V.le F. D'alcontres, 31, I-98166, Messina, Italy.

**Abstract**

A storage scheme based on racetrack memory, where the information can be coded in a domain or a skyrmion, seems to be an alternative to conventional hard disk drive for high density storage. Here, we perform a full micromagnetic study of the performance of synthetic antiferromagnetic (SAF) racetrack memory in terms of velocity and sensitivity to defects by using experimental parameters. We find that to stabilize a SAF skyrmion, the Dzyaloshinskii-Moriya Interaction in the top and the bottom ferromagnet should have an opposite sign. The velocity of SAF skyrmion and SAF Néel domain wall are of the same order and can reach values larger than 1200 m/s if a spin-orbit torque from the spin-Hall effect with opposite sign is applied to both ferromagnets.

# I. INTRODUCTION

The advent of a scenario where Cloud Computing and Internet of Things are merged together is pushing the research efforts to find a technology that can go beyond the complementary metal-oxide semiconductor (CMOS), especially in terms of storage and computing [1]. Spintronics, with its subfields, is a promising candidate for that, owed to the possibility to use the degree of freedom of the spin angular momentum in addition to the charge of the electrons. Concerning the storage, two different categories of devices are promising: spin-transfer-torque-MRAMs (STT-MRAMs) [1–5] and racetrack memories [6–9]. While STT-MRAMs (already on the market) can serve as "universal memory" [1] (low writing energy, high read speed, and ideally infinite endurance) [2,5], racetrack memories can be used as worthy alternative to hard disk drives in storage memory, with the advantage that no mechanical parts are necessary [9].

Since the first domain wall based racetrack memory presented in Ref. [6], performance improvements have been achieved by using materials with perpendicular anisotropy [10,11], spin-orbit interactions, such as spin-Hall effect (SHE) [12,13] and interfacial Dzyaloshinskii-Moriya Interaction (IDMI)) [7,8,14,15], and interlayer exchange coupling (IEC) [16,17]. All previous ingredients gave rise to a racetrack memory where the perpendicular anisotropy and the IEC, together with the IDMI, stabilize DWs in synthetic antiferromagnets (SAFs) while SHE drives the DW motion at a velocity near 800 m/s [17].

Recently, an alternative scheme of racetrack memory based on skyrmions has been proposed in Refs. [18–21]. Skyrmions are topologically protected magnetic configurations that can be used for different applications [22–25]. In particular, in racetrack memories they can directly code the information, i.e. the presence/absence of the skyrmion represents the bit "1"/"0" [18]. Skyrmion motion in ultrathin ferromagnetic materials has been demonstrated experimentally in extended Ta/CoFeB/MgO multilayers [26], and in Pt/CoFeB/MgO [27], where a skyrmion velocity near 120 m/s has been measured. The main technological challenge to be overcome is the experimental control of a single skyrmion nucleation by an electrical current and its electrical detection [28,29]. On the other hand, the main fundamental limitations of the skyrmion based racetrack memories are (i) the skyrmion Hall effect (two velocity components, parallel and perpendicular to the electrical current direction) [21,30] (ii) a transient breathing mode [21] and (iii) velocities well below the ones of DWs for a fixed set of physical and geometrical parameters [18]. Lately, it has been shown that these issues could be solved by using a SAF skyrmion [31], opening a path for a more competitive skyrmion based racetrack memory. In that work [31], the tunnelling interlayer coupling was used to stabilize the SAF skyrmion [32,33].

Here, we consider the same experimental framework by Yang *et al.* [17] where the IEC is mediated by itinerant electrons via Ruderman-Kittel-Kasuya-Yosida (RKKY) field [34], resulting in an antiferromagnetic alignment of the two ferromagnets (other forms of IEC exist, such as Orange-Peel coupling, tunnelling etc.). Firstly, we benchmark our computations with the experimental data on DWs, and then we predict the performance of SAF skyrmion based racetrack by considering realistic parameters, including edge roughness and bulk defects in the form of disordered grains with random uniaxial perpendicular anisotropy. Our key finding is that skyrmions in a SAF racetrack have velocities approaching the ones of DWs. We also show that a skyrmion based SAF racetrack is less sensitive to edge roughness – an unavoidable type of defects in narrow nanotracks – than their DW counterpart, as long as the skyrmion motion occurs far from the edges. On the other hand, the presence of a disordered anisotropy, considering different size of the grains, introduces a pinning of both SAF DW and skyrmion, which is stronger when the average size of the grains is larger than the skyrmion diameter. In particular, this result shows that, when the grain size (GS) is larger than the skyrmion diameter, the pinning of the skyrmion originates only at the interface between two adjacent grains.

## II. MICROMAGNETIC MODEL

We perform micromagnetic simulations of multilayered nanowires similar to the ones proposed in Ref. [17]. They are composed of a 3 nm thick Platinum heavy metal (HM) (lower HM) with on top two perpendicular CoNi ferromagnetic layer (FMs) separated by a thin Ruthenium (Ru) layer designed to provide an antiferromagnetic exchange coupling [34] (see Fig. 1(a)) and a second HM on top of the whole stack (upper HM).

The nanowire is 100 nm wide and 1000 nm long and the thickness of both ferromagnets and Ru layer is 0.8 nm. The physical parameters of CoNi layers taken from [16,17], and equal for both ferromagnets, are: saturation magnetization $M_s$=600 kA/m, exchange constant $A$=20 pJ/m, uniaxial perpendicular anisotropy constant $k_u$=0.6 MJ/m$^3$, and damping $\alpha_G$=0.1 [16,17]. The interlayer exchange coupling constant $A^{ex}$ is fixed to -5.0x10$^4$ J/m$^2$ [16,17]. We use a discretization cell of 4x4x0.8nm$^3$, and introduce a Cartesian coordinate system with the *x*-, *y*- and *z*- axes lying along the length, the width and the thickness of the wire, respectively (see Fig. 1(a)). The numerical study is carried out by means of a self-implemented micromagnetic solver [35,36] that includes the SHE, IDMI and IEC.

The total micromagnetic energy density of the system under investigation is (the superscripts *L* and *U* refer to lower and upper FMs):

$$\begin{aligned}\varepsilon_{tot}^{L,U} =\ &A(\nabla\cdot\mathbf{m})^2 + \varepsilon^{ex} \\ &+ D^{L,U}\left[m_z^{L,U}\nabla\cdot\mathbf{m}^{L,U} - (\mathbf{m}^{L,U}\cdot\nabla)m_z^{L,U}\right] \\ &- k_u\left(\mathbf{m}^{L,U}\cdot\hat{z}\right)^2 - \frac{1}{2}\mu_0 M_s \mathbf{m}^{L,U}\cdot\mathbf{H}_m^{L,U}\end{aligned} \qquad (1)$$

where $m_x$, $m_y$ and $m_z$ are the x-, y-, and z-components of the normalized magnetization **m**, respectively. The interlayer exchange energy is given by $\varepsilon^{ex} = -\frac{A^{ex}}{t_{RU}}(\mathbf{m}^L\cdot\mathbf{m}^U)$ (same contribution for both FMs), where $t_{RU}$ =0.8 nm is the thickness of the Ru layer [17]. $D$ is the parameter taking into account the intensity of the IDMI. More specifically, $D^L$ ($D^U$) refers to the lower (upper) FM. According to our definition, $D^L$ and $D^U$ are materials properties (they are not related to the multilayer geometry), i.e. in the specific case where both lower (HM/FM) and upper (FM/HM) interfaces are the same, one would have $D^L=D^U$.

The Pt/FM interface produces the lower IDMI that we have swept from 0 mJ/m² to 4.0 mJ/m², while the upper IDMI derives mainly from an upper HM. $\hat{z}$ is the unit vector along the z-direction, $\mu_0$ is the vacuum permeability. $\mathbf{H}_m$ is the magnetostatic dipolar field, which is computed by considering both ferromagnetic layers. The boundary conditions related to the IDMI are $\frac{d\mathbf{m}^{L,U}}{dn^{L,U}} = \frac{1}{\xi^{L,U}}(\hat{z}\times\mathbf{n}^{L,U})\times\mathbf{m}^{L,U}$ [21,37,38], where **n** is the unit vector normal to the surface and $\xi^{L,U} = \frac{2A}{D^{L,U}}$ is a characteristic length.

In the rest of the paper, the names DW and skyrmion refer to SAF configuration unless otherwise specified. Moreover, we wish to underline that, when considering the dynamical analysis, the presence of two HMs gives rise to a lower and upper SHE when an electric current is passed through them (the electrical current here denotes a flow of electrons). In order to model this behavior, it is necessary to introduce two spin-Hall angles, $\theta_{SH}^L$ and $\theta_{SH}^U$, acting on the lower and upper FMs, respectively. As for $D^L$ and $D^U$, according to our definition, the two spin-Hall angles are only linked to material properties.

### III. STABILITY PHASE DIAGRAMS

We firstly evaluate the stability phase diagram for both DW and skyrmion magnetic states as a function of the IDMI in the lower FM ($D^L$) and the IDMI ratio (-$D^U/D^L$) with the aim to find a region where both states are stable.

Fig. 1(b) shows these results, where we considered, as initial state, a left-handed Néel DW for DW and Néel (hedgehog-like) skyrmion for the skyrmion. We display the region $-D^U/D^L>0$ being the only region of the phase diagram where skyrmion can be stabilized. This is the first important qualitative result of our study, i.e. the IDMI of the two HMs at the two HM/FM and FM/HM have to be opposite and then the upper and lower HMs should be *different* (see Appendix A). In our case, because the lower HM is Pt, the upper HM should be, for instance, W.

The phase diagram for the DW shows that a stable Néel DW (see snapshot Fig. 1(c)) is achieved for the whole range of parameters analyzed. For the skyrmion, three different regions can be identified. The first one is characterized by the uniform antiferromagnetic state (AF), where the IDMI is not large enough to stabilize the skyrmion. The second region involves hedgehog-like or Néel skyrmion (Skx), which is achieved when $D^L$ is large enough ($D^L \geq 2.3$ mJ/m$^2$). This threshold IDMI for the skyrmions stabilization is the first important difference with respect to the DW. The skyrmion core in the lower FM is oriented in the negative *z*-direction, with its DW spins pointing outward and the surrounded magnetization is along the positive *z*-direction. The upper skyrmion has an opposite orientation of the spins (core along the positive *z*-axis with inward DW spins at the transition region) and the magnetization pointing downward (-*z*-axis) (see snapshot in Fig. 1(c)). At $D^L=4.0$ mJ/m$^2$, the minimum IDMI ratio ($-D^U/D^L$) to stabilize skyrmions is 0.2 mJ/m$^2$. In a previous paper [31], the skyrmion was stabilized without an upper HM ($D^L$ was set equal to 3.5 mJ/m$^2$) because of different physical parameters.

The third region, obtained at very high $D^L$ (>3.2 mJ/m$^2$), shows the stabilization of an extended skyrmion (ES), which is widened over the whole length of the strip. This state is linked to the bimeron instability (see snapshot of Fig. 1(c)), as well explained in Ref. [39], and its size is related to the confining potential given by both the magnetostatic energy and the IDMI boundary conditions.

In summary, the comparison between the two diagrams leads to the following main results:
- Skyrmion is obtained only in a narrow range of $D^L>0$ and positive IDMI ratio ($-D^U/D^L>0$), while DW is stable for all the values. The stability of a skyrmion is also achieved for $D^L<0$ and $-D^U/D^L>0$ (not shown) as can be understood for symmetric reasons.
- A region where both skyrmion and DW are stable exists (cyan region in Fig. 1(b)), enabling their comparison by using the same set of parameters. In particular, for the dynamical analyses, we will focus on two working points 'A' ($D^L=2.5$ mJ/m$^2$, $-D^U/D^L=1.0$) and 'B' ($D^L=3.5$ mJ/m$^2$, $-D^U/D^L=0.4$) as indicated directly in Fig. 1(b);

- In those regions, the presence of either skyrmion or DW depends on the nucleation process.

## IV. SAF DW MOTION

The first step of the dynamical study is the analysis of a SAF racetrack similar to the one proposed in Ref. [17], where the lower ferromagnet is coupled to a Pt HM, and no upper HM is considered. We use *state-of-the-art* values of $D^L$=2.5 mJ/m$^2$ [40–42] and consider three different IDMI ratios -$D^U$/$D^L$: 0.0, 0.4, and 1.0 (in Ref. [17], the authors claim that there might be an upper IDMI originated from the lower HM/FM interface and/or the Ru layer). We also take into account the effect of the edge roughness (two different patterns for the lower and upper FMs) as computed by the algorithm developed in Ref. [43]. In detail, they are obtained by randomly removing regions from the strip edges with a uniform probability distribution characterized by the typical roughness size $D_G$. In our study, we consider a $D_G$=12 nm [44] averaging the results of the velocity over five different random patterns. The electrical current $j_{HM}$ is injected along the *x*-axis via the lower HM in order to originate the SHE only for the lower FM, with $\theta_{SH}^L$=0.12 [45].

Fig. 2 summarizes the results of this study, where the curve with blue stars in Fig. 2(a) (experimental data from Ref. [17]), has been chosen as reference. The lower and upper DWs move tied along the *x*-direction because of the IEC (see Supplemental MOVIES 1 and 2 for $j_{HM}$=1.50 and 3.25x10$^8$ A/cm$^2$, respectively). The velocity-current relation (*v* indicates the velocity along the *x*-axis) is independent of the IDMI ratio, such that the three numerical curves are almost overlapped. The pinning due to the edge roughness (see Supplemental MOVIE 3 for $j_{HM}$=0.50x10$^8$ A/cm$^2$) gives rise to a threshold current to move the DW ($|j_{HM}|$>0.75x10$^8$A/cm$^2$). As can be observed in the previous movie, the DW motion is characterized by a spatial tilting of its profile normal. The tilting angle of the DW spatial profile reverses symmetrically with the sign of the current (see also for comparison Fig. 1b in [17]), and it is robust against the edge roughness (compare Supplemental MOVIES 1 and 4, the latter obtained for a perfect SAF and $j_{HM}$=-1.50x10$^8$ A/cm$^2$). Our simulations point out that the DW tilting is essentially due to the IDMI boundary conditions [46]. In fact, when the IDMI boundary conditions are neglected and regular exchange boundary conditions are used, the DW profile normal keeps straight pointing along the *x*-axis (see Supplemental MOVIE 5 for a perfect SAF without IDMI boundary conditions and $j_{HM}$=-1.50x10$^8$ A/cm$^2$). In addition, as the IDMI ratio increases, the tilting angle reduces for a fixed value of the current (see Supplemental MOVIE 6 for a perfect SAF, $j_{HM}$=-1.50x10$^8$ A/cm$^2$ and -$D^U$/$D^L$=1.0). In other words, the larger is the current, the larger is the tilting. From this analysis to benchmark our model, we can conclude that the numerical outcomes are in good agreement with the experimental findings and the numerical model can be used for the predictions of dynamical proprieties of DW and Skyrmion is similar systems.

In the rest of the paper, we focus on the parameters related to the working points 'A' and 'B' of Fig. 1(b). First of all, we study in 'A' ($\theta_{SH}^{L} = 0.12$) the dependence of the DW velocity-current relation as a function of $\theta_{SH}^{U}$.

For $\theta_{SH}^{U} = 0$, the velocity-current relation is linear and the magnitude of the maximum velocity is 955 m/s for $j_{HM}$~3.25x10$^8$ A/cm$^2$ (Fig. 3(a) red squares). The key reason of such large velocity is a trade-off between the IEC torque, that tends to maintain the Néel DW configuration, and the SHE torque, that tends to rotate the DW inner magnetization towards a Bloch state. In other words, the large IEC allows for the injection of larger currents before reaching the saturation regime as well explained in Ref. [17]. As expected, in the absence of the edge roughness and/or other type of defects no threshold current to move the DW is observed.

The effect of the $\theta_{SH}^{U}$ is summarized in Fig. 3(a). While for negative $\theta_{SH}^{U}$ (-0.06 and -0.12 black circles, and olive up-triangles, respectively) the velocities grow proportionally, for positive values of $\theta_{SH}^{U}$ (blue down-triangles, $\theta_{SH}^{U} = 0.06$), they decrease. In this case, the upper SHE acts as a "brake" in slowing down the DW velocity. If $\theta_{SH}^{U}$ is exactly the same (0.12, not shown), the two SHE torques exactly balance each other and DW does not move. The non-linear behavior observed for $\theta_{SH}^{U}$=-0.12 at high current ($|j_{HM}|$>2.00x10$^8$ A/cm$^2$) is ascribed to the tilting of the DW inner magnetization towards the spin-polarization direction (*y*-axis), leading to a reduction of the effective SHE torque acting on the DW.

This analysis leads to the second key result of this paper: the SHE angle of the upper HM should have a sign opposite to the lower HM one to enhance the velocity performance. To this aim, can use the W (see Appendix B). Fig. 3(b) shows the velocity-current relation related to the working point 'B', compared with the one of 'A'. No significant changes are observed.

In order to provide a more realistic analysis, we also introduce a random variation of the perpendicular anisotropy constant $k_u$ (disordered anisotropy) to simulate the presence of disorder (i.e. crystalline grains, defects, local variation of heavy metal and/or ferromagnet thickness can be accounted by these random variation of the PMA constant). In particular, we introduce a $k_u$ dispersion of $\pm$10% around its mean value of 0.80 MJ/m$^3$, and we create grains by means of the Voronoi tessellation algorithm. Three different patterns are studied, where the grain size has been determined in comparison with the skyrmion size. Specifically, the skyrmion diameter is 2R$_{Sk}$=40 nm, therefore GS<2R$_{Sk}$, GS=2R$_{Sk}$, and GS>2R$_{Sk}$ refer to grain sizes smaller, equal and larger than the skyrmion diameter: GS=13 nm, 40 nm, and 63 nm, respectively. Two different grain patterns have been considered for the upper and the lower FMs with a fixed mean grain size GS.

Fig. 3(c) shows the results of the effect of the disordered anisotropy on DW motion in a strip with no edge roughness. The presence of grains induces an asymmetric DW pinning, which is higher for the largest GS considered here ($j_{HM}$=-1.25x10$^8$ A/cm$^2$). However, this asymmetric DW pinning is sample-to-sample dependent and this asymmetry is strongly reduced when the results are averaged over a large number of simulations for different grain patterns with the same characteristic size. Similar results are obtained when the disordered anisotropy is introduced in a rough strip (see Fig. 3(d)), where the combined effect of grains and edge roughness results in an increase of the depinning threshold current.

**V. SAF SKYRMION MOTION**

Considering the working points 'A' and 'B', the current-driven skyrmion dynamics is also studied. The two skyrmions in the upper and the lower FMs move tied and do not exhibit either the skyrmion Hall effect (i.e. the component of the velocity (*y*-axis) perpendicular to the electrical current direction (*x*-axis) is zero [31]), or the transient breathing mode observed in Ref. [21]. Within a collective approach, the total skyrmion number *S* is zero, as already described in Ref. [31]. Therefore, in the Thiele's equation [21,47], the "gyrocoupling vector" $\mathbf{G} = (0,0,4\pi S)$ is zero [21]. The analysis of the effect of the spin-Hall angle leads to similar results as for the DW dynamics in term of influence of $\theta_{SH}^{U}$ (Fig. 4(a) and (b)). On the other hand, Fig. 4(c) shows that the skyrmion motion is not affected by the presence of the edge roughness as expected (see Supplemental MOVIE 7 to compare the skyrmion motion in the lower FM of a perfect and rough track for $j_{HM}$=1.50x10$^8$ A/cm$^2$). This result makes SAF skyrmion racetrack more promising as compared to the one based on a single HM/FM bilayer, where the presence of the skyrmion Hall angle leads inevitably single skyrmion to interact with the edge roughness for high currents. Moreover, we analyze the effect of disordered anisotropy, as for DW motion. Also for the skyrmion, the disordered anisotropy gives rise to an asymmetric pinning that depends on the spatial distribution of the grains, and it is higher for the largest disordered anisotropy size (see Figs 3(d)). This aspect is the key ingredient that originates a threshold current necessary for the skyrmion motion. In addition, those results point out that the disordered anisotropy is a fundamental feature to be taken into account when scaling the skyrmion diameter. Finally, the combined effect of disordered anisotropy and edge roughness does not significantly affect the velocity-current relations in the SAF heterostructure (not shown).

Some differences with recent works studying the skyrmion motion in SAF [31] and in antiferromagnetically-coupled structure [48] are briefly discussed below. Here, we have set the parameters of our computations after a benchmark with experimental data of the SAF DW dynamics. Ref. [31] deals with an interlayer exchange coupling originated from spin-tunneling (the two ferromagnets are separated by an insulator). In Ref. [48], a real antiferromagnetic skyrmion motion is driven by the spin-transfer torque (STT) of the Zhang-Li form [49] due to an in-plane current. In ultrathin systems, where the ferromagnetic thickness is less than 1 nm, it is very likely that the torque is mainly due to SHE (anti-damping torque of Slonczewski type) rather than STT. Indeed, due to the very thin thickness of the ferromagnets, most of the current flows through the heavy metals, so that the STT is usually negligible. In addition, the geometry of STT and SHE is profoundly different since the STT only cares about the magnetization gradient, while the SHE is uniquely sensitive to the chirality of the skyrmion (Néel, Bloch etc.). A final remark is that the key drawback for antiferromagnetic solitons is their electrical detection.

## VI. COMPARISON BETWEEN SAF DW AND SKYRMION MOTION

We study, as reference, a racetrack made by a single HM/FM bilayer. The geometrical and physical parameters are: strip length along *y*-direction, width along *x*-direction and thickness 1000 nm, 100 nm, and 1 nm, respectively, $M_s$=1000 kA/m, $A$=20 pJ/m, $k_u$=0.8 MJ/m$^3$, $\alpha_G$=0.015, $D$=1.8 mJ/m$^2$, $\theta_{SH} = -0.33$ (same as Scenario B of Ref. [21]). Fig. 5(a) shows the results of this study, highlighting that single DW is faster than single skyrmion for the whole current range, with a velocity difference of about 300 m/s for |$j_{HM}$|>30 MA/cm$^2$. Similar results are also observed for other parameters.

The last part of the paper is focused on a comparison between DW and skyrmion motion. Let's consider the working points 'A' ($D^L$=2.5 mJ/m$^2$,-$D^U$/$D^L$=1.0, see Fig. 5(b)) and 'B' ($D^L$=3.5 mJ/m$^2$, -$D^U$/$D^L$=0.4, see Fig. 5(c)), where only the SHE in the lower FM is taken into account. In a perfect strip, the DW velocity is larger than the skyrmion one.

Now, let's consider a strip with edge roughness. Fig. 5(d) shows the results for the working point 'A'. At low currents, when the DW remains pinned, skyrmion is faster (see Supplemental MOVIE 8 for the comparison of DW and skyrmion motion in the lower FM of a rough track, for $j_{HM}$=0.50x10$^8$ A/cm$^2$), whereas, for current values up to 2.00x10$^8$ A/cm$^2$, their velocities are comparable. When |$j_{HM}$|>2.00x10$^8$ A/cm$^2$, the DW velocity is larger than the one of the skyrmion with a difference smaller than 200 m/s (see Supplemental MOVIE 9 for the comparison of DW and skyrmion motion in the lower FM of a rough track, for $j_{HM}$=3.25x10$^8$ A/cm$^2$).

Finally, we compare DW and skyrmion motion in presence of disordered anisotropy (GS=13nm, GS=40nm=$2R_{Sk}$, and GS=63nm). The disordered anisotropy introduces pinning for both DW and skyrmion dynamics, and the threshold current increases as the GS enlarges (compare Figs 3(d) and 4(d)). Therefore, here we present the results for the worst case only, i.e. when the grain size is larger than the skyrmion diameter (GS=63nm>$2R_{Sk}$).

Fig. 6(a) reports the comparison considering the disordered anisotropy only. The skyrmion exhibits a slightly larger depinning current and a smaller velocity than the DW one in the whole current range. This result demonstrates that the skyrmion motion is more sensitive to disordered anisotropy than DW. Fig. 6(b) shows the combined effect of disordered anisotropy and edge roughness. DW and skyrmion exhibit the same depinning current and a comparable velocity up to $|j_{HM}|=2.00 \times 10^8$ A/cm$^2$ (see Supplemental MOVIE 10 for the comparison of DW and skyrmion motion in the lower FM of a rough track with disordered anisotropy, for $j_{HM}=1.75 \times 10^8$ A/cm$^2$). At larger currents, the DW is still faster than skyrmion (see Supplemental MOVIE 11 for the comparison of DW and skyrmion motion in the lower FM of a rough track with disordered anisotropy, for $j_{HM}=3.00 \times 10^8$ A/cm$^2$).

The results of the comparison are summarized in Figs. 6(c)-(d) where the ratio between DW and skyrmion velocity in a SAF racetrack (Fig. 6(c))) and in a single HM/FM bilayer (Fig. 6(d)) are shown. While in the latter, this ratio ranges between 3 to 8, in SAF tracks, it is reduced in the range 1.35 to 1.55, proving that the velocity of skyrmions in SAF racetracks is much closer to the one of DWs. This is the final quantitative important result of our work.

## VII. SUMMARY AND CONCLUSIONS

In summary, we have studied a SAF racetrack based on both DW and skyrmion. The static analysis has pointed out that, while Néel DW can be stabilized in all the range of positive IDMI ratio (including the case of zero upper IDMI), skyrmion can be obtained only if the upper IDMI is large enough. This result entails that, by considering state-of-the-art parameters, it is necessary to couple the upper ferromagnet to an upper HM, which has to be characterized by an IDMI parameter sign opposite to the one of the lower HM.

For the motion driven by SHE, the presence of an upper HM introduces an additional degree of freedom for controlling both DW and skyrmion motion, allowing one to increase/reduce the velocity if the two spin-Hall angles have different/same sign. The comparison of DW and skyrmion in a track with edge roughness and disordered anisotropy has shown that (i) skyrmion velocity can approach the DW one in SAF racetrack, (ii) a SAF skyrmion, being not subject to the skyrmion Hall effect, is insensitive to the edge roughness, and (iii) the disordered anisotropy introduces pinning for both DW and skyrmion motion, and the threshold current increases as the grain size gets larger. In addition, racetrack based on SAF Skyrmion/DW can be more scalable in term of distance between two adjacent strips as compared with the standard racetracks based on single HM/FM bilayer. This is because the dipole interactions of the two opposite ferromagnetic layers partially cancel each other, thus inducing reduced disturbance as the FM strips becomes closer.

Finally, the use of SAF racetrack has a technical advantage over real antiferromagnetic racetrack: it is possible to achieve electrical detection of skyrmion by means of a magnetic tunnel junction (MTJ) built on top of the racetrack (a variation of the magnetic state below the MTJ is linked to a change in the tunneling magnetoresistive signal).

## Acknowledgments

The authors acknowledge the executive programme of scientific and technological cooperation between Italy and China for the years 2016-2018 (code CN16GR09) title "Nanoscale broadband spin-transfer-torque microwave detector" funded by Ministero degli Affari Esteri e della Cooperazione Internazionale. R. T. and M. R. also acknowledge Fondazione Carit - Projects – "Sistemi Phased-Array Ultrasonori", and "Sensori Spintronici". The work by E. M. was supported by project WALL, FP7-PEOPLE-2013-ITN 608031 from European Commission, project MAT2014-52477-C5-4-P from Spanish government, and project SA282U14 from Junta de Castilla y Leon. A. M. acknowledges support from the King Abdullah University of Science and Technology (KAUST). The authors acknowledge André Thiaville for very useful discussions.

## APPENDIX A: IDMI sign for the two heavy metals

In order to stabilize the SAF configuration, the two heavy metals should be characterized by an *opposite* sign of the IDMI parameter. In fact, if we consider the SAF configuration for the UP and DOWN domains, and the same sign of the IDMI parameter for both heavy metals ($D^U=D^L$), we obtain an opposite chirality of both DW and skyrmion in the lower and upper ferromagnets (see Fig. A1(a)). On the other hand, if we consider an opposite sign of the IDMI parameter ($D^U=-D^L$), we achieve the SAF configuration of both DW and skyrmion (see Fig. A1(b)), where both magnetic patterns are characterized by the same left-handed chirality. This is the reason why we have considered a negative IDMI ratio in Fig. 1(b) of the main text. We can conclude that, if, for instance, the lower HM is Platinum, the upper one should be for instance Tungsten.

## APPENDIX B: spin-Hall angle for the two heavy metals

In order to enhance the velocity performance of SAF DW and skyrmion, the two heavy metals should be characterized by an *opposite* spin-Hall angle. In fact, if we consider that the two heavy metals have the same spin-Hall angle, i.e. the current flowing in the lower layer induces a positive spin-polarization (dots accumulating at the bottom interface), while the current flowing in the upper layer creates a negative spin-polarization (crosses accumulating at the bottom interface), the lower and upper DWs move in opposite directions (see Fig. A2(a)), leading to a reduction of the velocity (see Fig. 3(a) of the main text). On the other hand, if the two spin-Hall angles have an opposite sign, the two DWs move in the same direction (see Fig. A2(b)), yielding an increase of the velocity (see Fig. 3(a) of the main text). The same implications are valid for skyrmions. We can conclude that, if, for instance, the lower HM is Platinum, the upper one should be Tantalum or Tungsten. Since the combination Platinum/Tungsten is suitable for the IDMI sign as well, this could be a possible double heavy metal SAF configuration.

FIG. 1. (a) Sketch of the SAF multilayer under investigation, where the ferromagnets (FM), separated by a Ru layer, are sandwiched between different heavy metals (HM). (b) Stability phase diagram for DW and skyrmion (the initial state is a left-handed Néel DW for DW and Néel skyrmion for the skyrmion) of the magnetization state as a function of the lower IDMI parameter $D^L$ and the IDMI ratio. The acronyms AF, DW, Skx, and ES mean AntiFerromagnetic state, Néel Domain Wall, Skyrmion, and Extended Skyrmion, respectively. A and B refer to the working points considered for the study of DW and skyrmion motion. (c) Snapshots representing examples of the spatial distribution of the magnetization for DW (bottom right), Skx (bottom left), and ES (top). A color scale, linked to the $z$-component of the magnetization, is also indicated.

FIG. 2. A comparison of the numerical velocity-current relation with the experimental one (blue stars in Ref. [17]). The computations concern a fixed $D^L$=2.5 mJ/m$^2$, and three different IDMI ratios, as indicated in the legend. The edge roughness is taken into account for both ferromagnetic layers by means of the algorithm developed in Ref. [43] by using $D_g$=12 nm. The results are averaged over 5 different edge roughness distributions.

FIG. 3. DW velocity-current relation in (a) working point 'A' of Fig. 1(b) ($D^L$=2.5 mJ/m$^2$, $-D^U/D^L$=1.0), when $\theta_{SH}^L$ is fixed to 0.12 and $\theta_{SH}^U$ changes as indicated in the main panel, and (b) working point 'B' of Fig. 1(b) ($D^L$=3.5 mJ/m$^2$, $-D^U/D^L$=0.4) compared with 'A'. DW velocity-current relation in working point 'A' when the disordered anisotropy is introduced in a strip (c) without and (d) with edge roughness. The skyrmion diameter is 2R$_{Sk}$=40 nm, therefore the notations GS<2R$_{Sk}$, GS =2R$_{Sk}$, and GS>2R$_{Sk}$ refer to an average grain size of 13 nm, 40 nm, and 63 nm, respectively.

FIG. 4. Skyrmion velocity-current relation in (a) working point 'A' of Fig. 1(b) ($D^L$=2.5 mJ/m$^2$, $-D^U/D^L$=1.0), when $\theta_{SH}^L$ is fixed to 0.12 and $\theta_{SH}^U$ is changed as indicated in the main panel, (b) working point 'B' of Fig. 1(b) ($D^L$=3.5 mJ/m$^2$, $-D^U/D^L$=0.4) compared with 'A', and (c) working point 'A', when $\theta_{SH}^L$ is fixed to 0.12 and $\theta_{SH}^U$=0, with (green stars) and without (red squares) edge roughness. (d) Skyrmion velocity-current relation in presence of disordered anisotropy compared with the case of racetrack without edge roughness (red squares). The notations GS<2R$_{Sk}$, GS=2R$_{Sk}$, and GS>2R$_{Sk}$ refer to an average grain size of 13 nm, 40 nm, and 63 nm, respectively.

FIG. 5. A comparison between the velocity-current relation of (a) SAF skyrmion and skyrmion in racetrack made by a HM/FM bilayer, (b) SAF DW and skyrmion in working point 'A' ($D^L$=2.5 mJ/m$^2$, $D^U/D^L$=1.0), (c) SAF DW and skyrmion in working point 'B' ($D^L$=3.5 mJ/m$^2$, $D^U/D^L$=0.4), and (d), SAF DW and skyrmion in working point 'A' including edge roughness. All the computations are performed for $\theta_{SH}^L$=0.12 and $\theta_{SH}^U$=0.00.

FIG. 6. A comparison between the velocity-current relation of DW and skyrmion in a strip with disorder anisotropy with grains larger than the skyrmion diameter (GS=63nm>2R$_{Sk}$), (a) without edge roughness, and (b) with edge roughness. (c) and (d) velocity ratio of DW and skyrmion in a SAF and single HM/FM bilayer racetrack respectively.

FIG. A1. Spatial distribution of the magnetization in lower (L) and upper (U) ferromagnet for both skyrmion and DW, when (a) $D^U$=$D^L$, and (b) $D^U$=$-D^L$.

FIG. A2. Sketch of SAF DW motion when the two heavy metals have (a) same and (b) opposite spin-Hall angle.

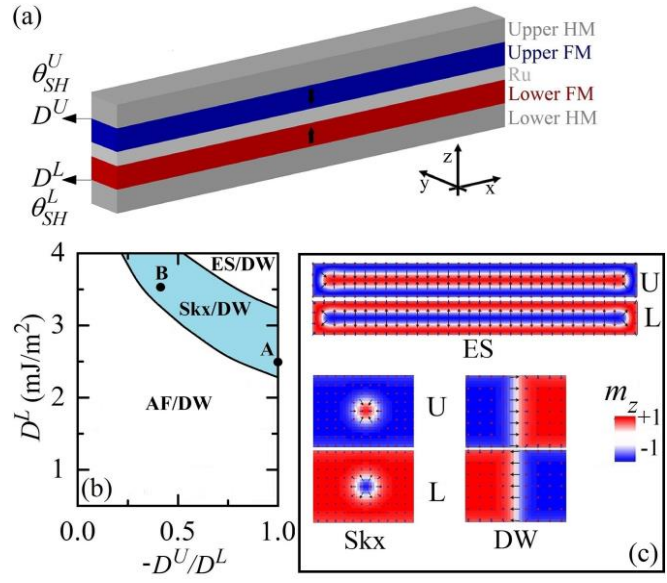

FIGURE 1

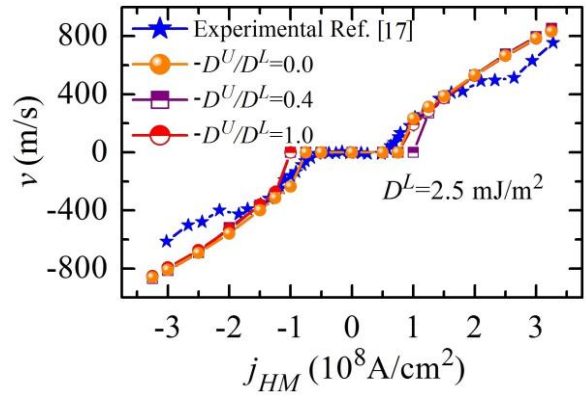

FIGURE 2

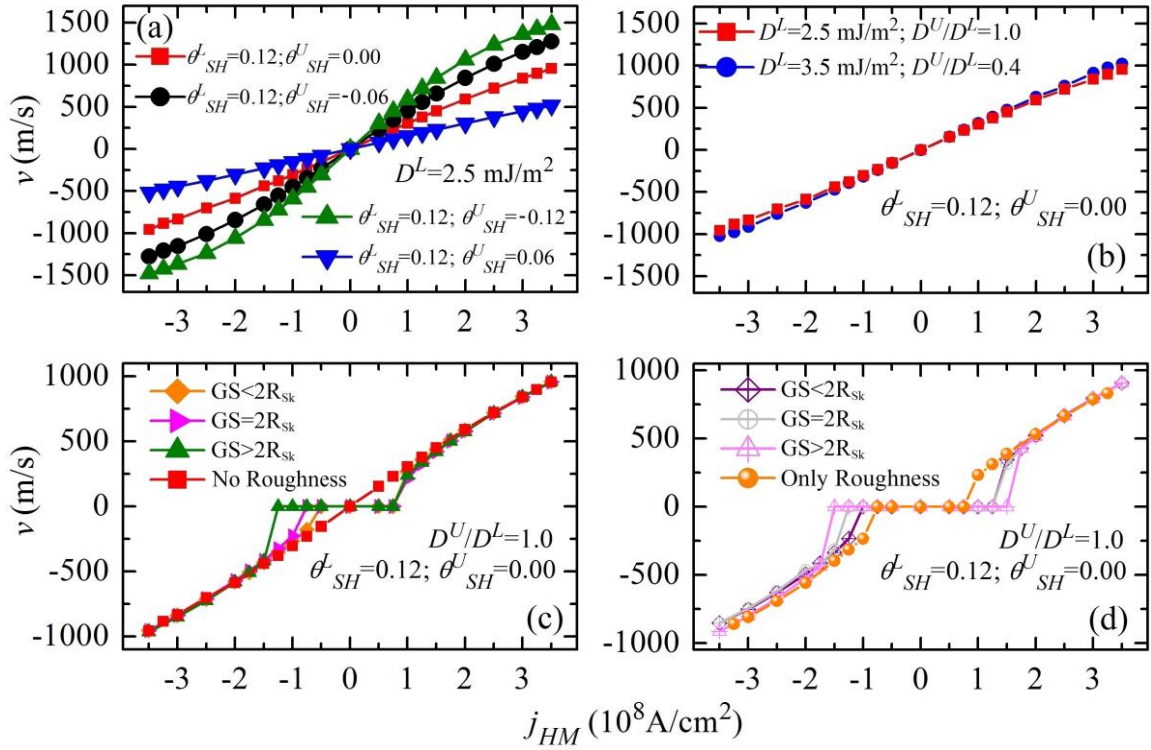

FIGURE 3

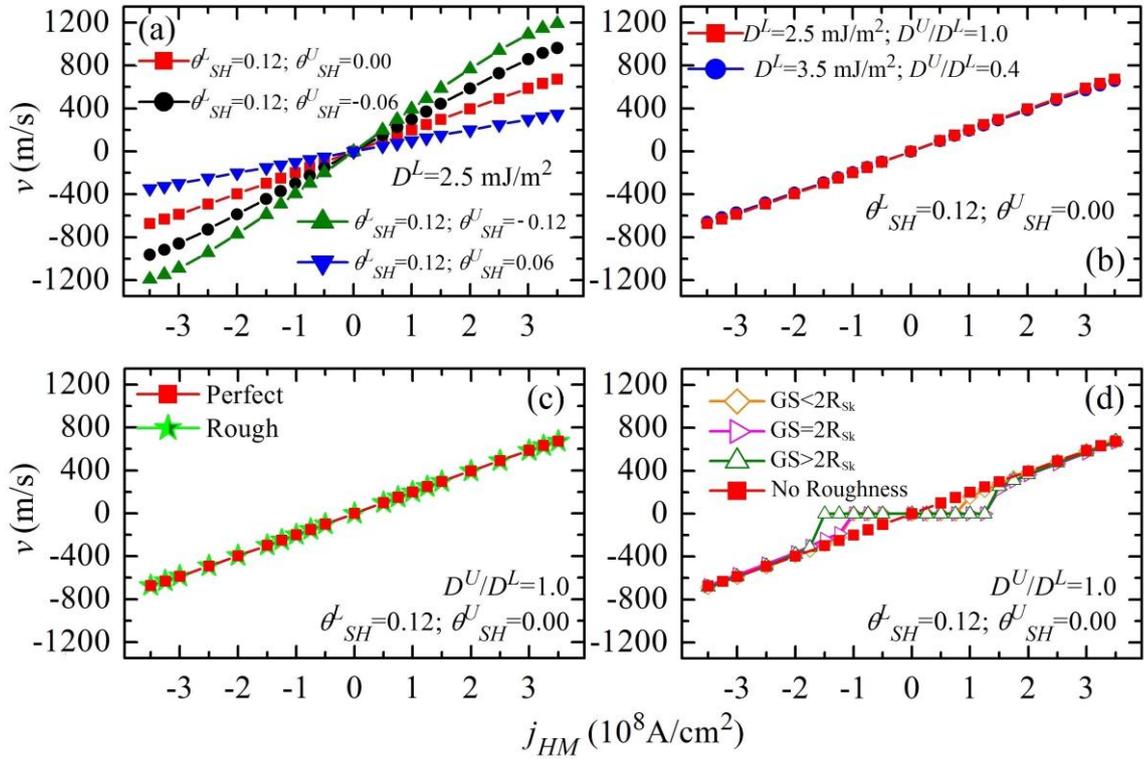

FIGURE 4

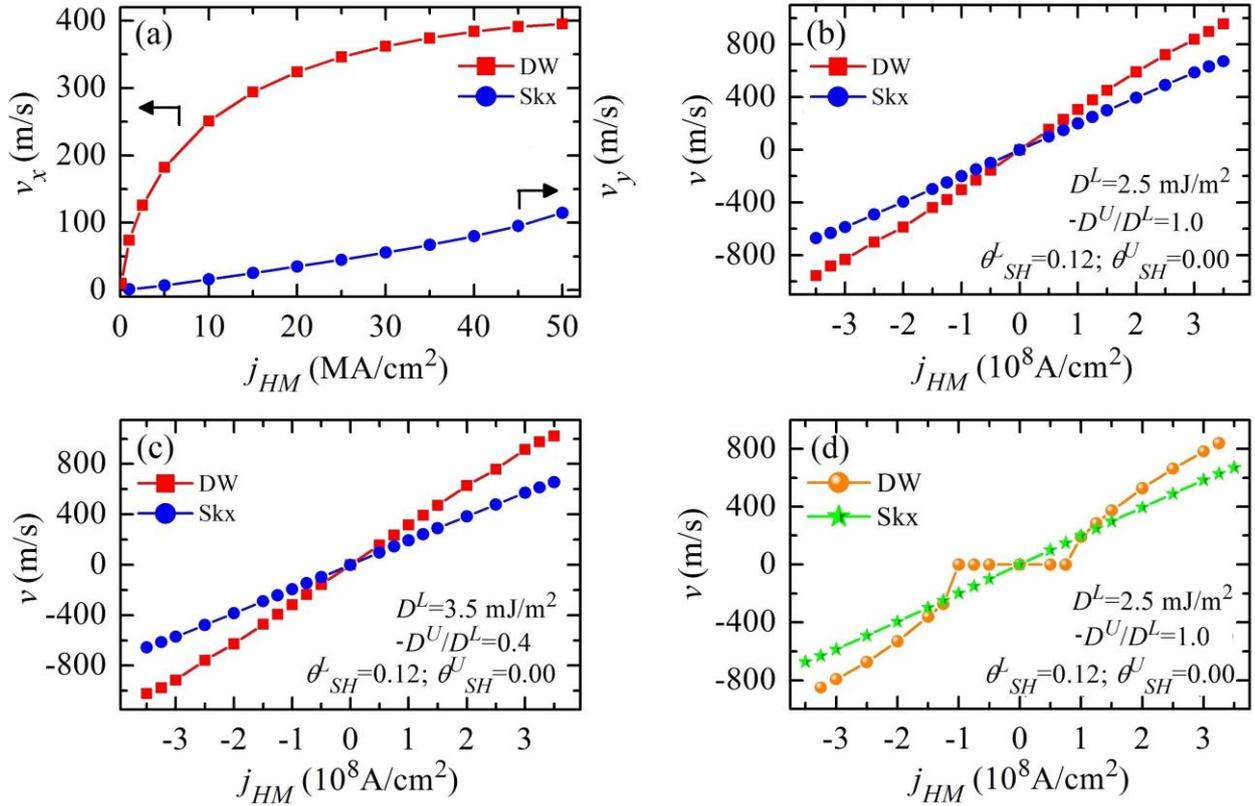

FIGURE 5

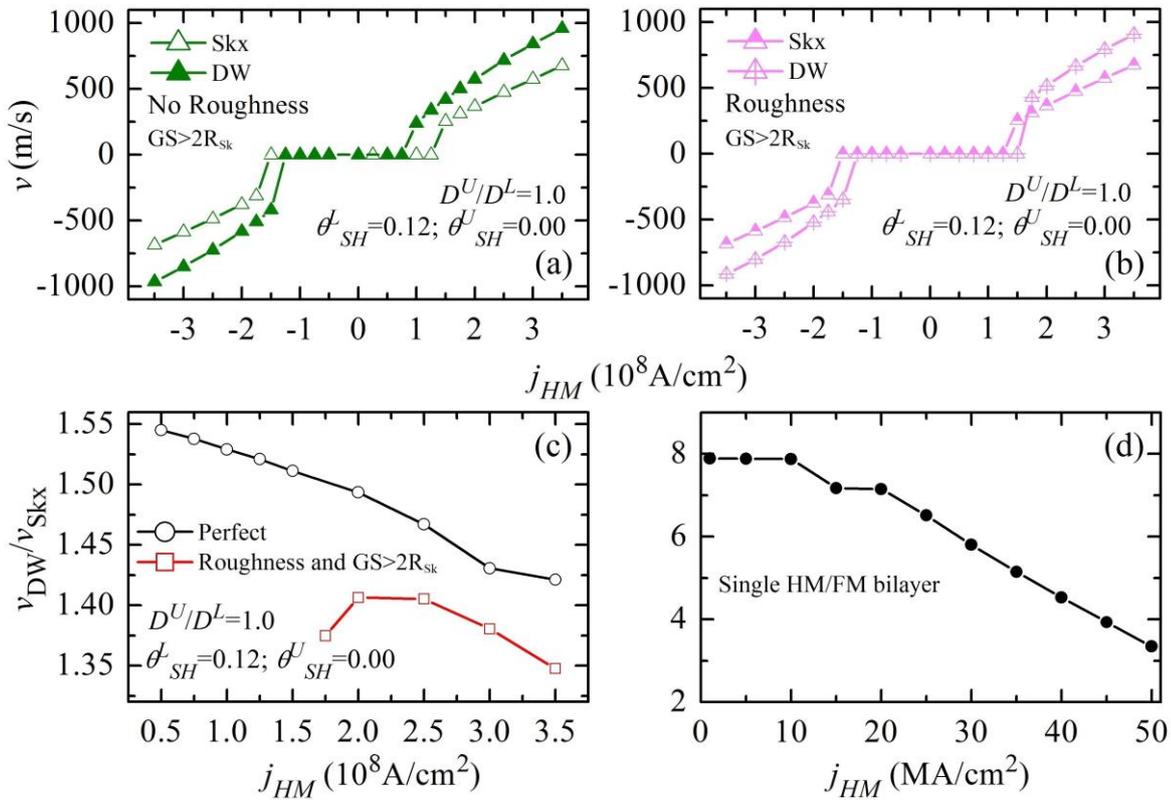

FIGURE 6

FIGURE A1

FIGURE A2